\documentclass[pra,twocolumn,superscriptaddress]{revtex4-2}

\usepackage[dvipsnames]{xcolor}
\usepackage{graphicx}
\usepackage{amsmath}
\usepackage{amsthm}
\usepackage{amssymb}
\usepackage{times}
\usepackage{braket}

\usepackage[T1]{fontenc}

\newcommand{\ave}[1]{\langle #1 \rangle}

\theoremstyle{definition}

\begin{document}

\title{Collisional open quantum dynamics with a generally correlated environment: \\ Exact solvability in tensor networks}

\author{Sergey N. Filippov}
\affiliation{Department of Mathematical Methods for Quantum
Technologies, Steklov Mathematical Institute of Russian Academy of
Sciences, Gubkina St. 8, Moscow 119991, Russia}

\author{Ilia A. Luchnikov}
\affiliation{Russian Quantum Center, Skolkovo, Moscow 143025,
Russia}

\begin{abstract}
Quantum collision models are receiving increasing attention as
they describe many nontrivial phenomena in dynamics of open
quantum systems. In a general scenario of both fundamental and
practical interest, a quantum system repeatedly interacts with
individual particles or modes forming a correlated and structured
reservoir; however, classical and quantum environment correlations
greatly complicate the calculation and interpretation of the
system dynamics. Here we propose an exact solution to this problem
based on the tensor network formalism. We find a natural Markovian
embedding for the system dynamics, where the role of an auxiliary
system is played by virtual indices of the network. The
constructed embedding is amenable to analytical treatment for a
number of timely problems like the system interaction with
two-photon wavepackets, structured photonic states, and
one-dimensional spin chains. We also derive a time-convolution
master equation and relate its memory kernel with the environment
correlation function, thus revealing a clear physical picture of
memory effects in the dynamics. The results advance tensor-network
methods in the fields of quantum optics and quantum transport.
\end{abstract}

\maketitle

\section{Introduction}

Multipartite quantum systems are notoriously difficult to study.
So is the open dynamics of a quantum system interacting with a
multipartite or multimode environment. The environment usually
consists of enormously many particles or modes, which makes it
almost impossible to track the exact dynamics of the system
density operator $\varrho_S(t)$. The exact treatment of the
problem is possible in some exceptional cases
only~\cite{caldeira-leggett-1983,hu-1992,palma-1996,vacchini-2010,ferialdi-2016,fang-2018,burgarth-2021},
whereas one usually has to resort to some physical approximations,
e.g., the weak system-environment coupling with a timescale
separation between the bath correlation and the system
relaxation~\cite{davies-1974,schaller-2008,benatti-2010,rivas-2017,trushechkin-2021}.
Another approach is based on a past-future independence for
environment degrees of freedom interacting with the
system~\cite{li-2018} --- the assumption that is naturally
fulfilled in a conventional collision model (also known as the
repeated interactions model) with uncorrelated environment
particles~\cite{rau-1963,scarani-2002,attal-2006,pellegrini-2009,grimmer-2016}.
The latter approach has received increasing attention in the
analysis of quantum nonequilibrium steady
states~\cite{karevski-2009,filip-2021,heineken-2021}, bipartite
and multipartite entanglement
generation~\cite{heineken-2021,daryanoosh-2018,cakmak-2019},
quantum thermodynamical analysis of
micromasers~\cite{strasberg-2017}, quantum
thermometry~\cite{seah-2019}, and simulation of open quantum
many-body dynamics~\cite{purkayastha-2021,cattaneo-2021}; see the
recent review papers on collision
models~\cite{ciccarello-2021,campbell-2021}.

\begin{figure}
\includegraphics[width=7.5cm]{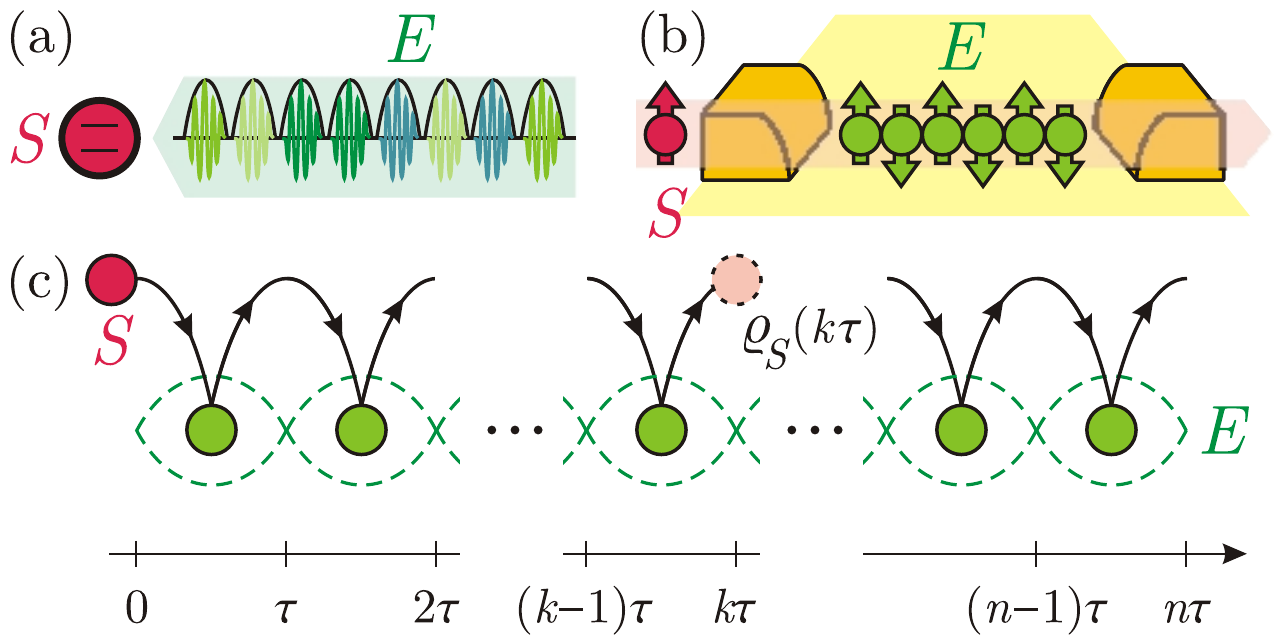}
\caption{(a) System interaction with an entangled multimode
environment state encoded in temporal modes of light. (b) Spin
transport through a one-dimensional chain. (c) Quantum collision
model with correlated environment.} \label{figure1}
\end{figure}

Collision models naturally emerge in time-bin quantum optics and
waveguide quantum electrodynamics, where the radiation field is
mapped into a stream of discrete time-bin modes of duration
$\tau$~\cite{pichler-2016,guimond-2017,ciccarello-2017,gross-2018,fisher-2018,cilluffo-2020,carmele-2020,ferreira-2021,wein-2021,maffei-2022}
that sequentially interact with the quantum system while the
radiation field propagates in space, see Fig.~\ref{figure1}(a).
However, in contrast to the conventional collision model with a
factorized environment, the radiation field represents a
correlated and structured environment that is difficult to deal
with even in the case of a single-photon
wavepacket~\cite{dabrowska-2020,dabrowska-2021}, not to mention
entangled multiphoton states generated from the cascade
emissions~\cite{jen-2017,cere-2018} or artificial photonic tensor
network
states~\cite{guimond-2017,dhand-2018,lubash-2018,istrati-2020,besse-2020,tiurev-2020,wei-2021}.
The latter ones are entangled multimode environment states
$\ket{\psi_E}$ encoded in temporal modes of light. The greater the
number $n$ of time bins the more complicated is the calculation of
the system density operator after $k$ collisions,
\begin{equation}
\varrho_S(k\tau) = {\rm tr}_{1,\ldots,k} \Big[ U_{Sk} \cdots
U_{S1} \varrho_S(0) \otimes \varrho_{1 \ldots k} \, U_{S1}^{\dag}
\cdots U_{Sk}^{\dag} \Big], \label{dynamics}
\end{equation}

\noindent where $\varrho_{1 \ldots k} = {\rm
tr}_{k+1,\ldots,n}[\ket{\psi_E}\bra{\psi_E}]$ is the reduced
density operator for $k$ environment modes (its dimension growing
exponentially with $k$) and $U_{Sk}$ is the evolution operator for
the system and the $k$-th mode. Exact and approximate solutions of
Eq.~\eqref{dynamics} are known for some exceptional correlated
environments and interactions~\cite{rybar-2012,filippov-2017};
however, a general solution to the system dynamics is still
missing.

The same computational problem emerges in a collision model for
spin transport through a chain of correlated atoms, e.g., a carbon
chain~\cite{zanolli-2010}, where the spin carrier moves
ballistically and sequentially interacts with correlated
environment particles, see Fig.~\ref{figure1}(b). The ground state
of a gapped one-dimensional local Hamiltonian for the spin chain
has a tensor network structure~\cite{dalzell-2019}. A seeming
complexity of the tensor network representation, as we show in
this paper, is in fact a key to an elegant solution to the
computational problem in Eq.~\eqref{dynamics}.

\section{Tensor network for the environment state}

\begin{figure}
\includegraphics[width=8.5cm]{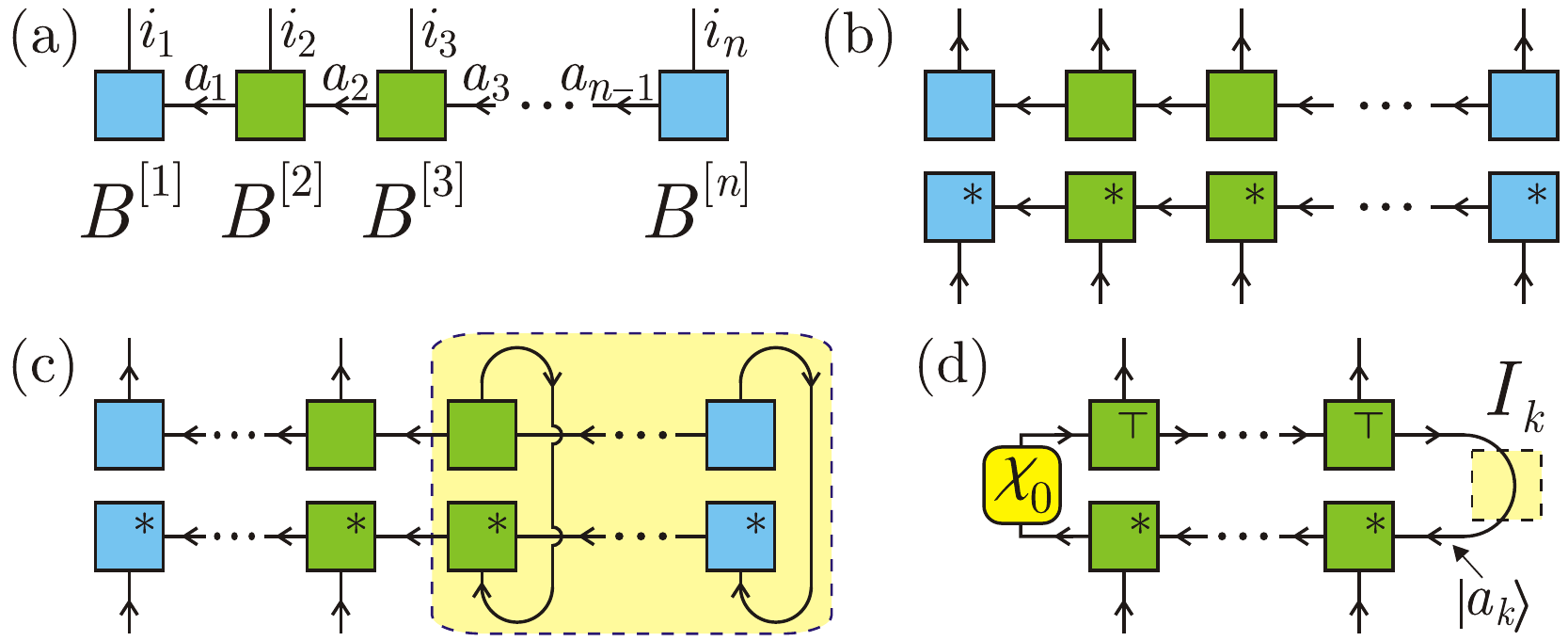}
\caption{Tensor diagrams for matrix product state $\ket{\psi_E}$
(a), density operator $\ket{\psi_E}\bra{\psi_E}$ (b), reduced
density operator (c) and its equivalent if environment has the
right-canonical form (d). The bond Hilbert space ${\cal
H}_{\text{bond} \# k} = \text{Span}(\{\ket{a_k}\})$. Outcoming and
incoming arrows stand for ket- and bra-components, respectively.}
\label{figure2}
\end{figure}

Any pure state of $n$ correlated $d$-dimensional particles adopts
the following form of a matrix product state
(MPS)~\cite{perez-garcia-2007,verstraete-2008,schollwock-2011,cirac-2021}:
\begin{equation}
\ket{\psi_E} = \sum_{i_1,i_2, \ldots, i_{n} = 0}^{d-1} B^{[1],i_1}
B^{[2],i_2} \cdots B^{[n],i_n} \ket{i_1 i_2 \ldots i_n},
\label{mps}
\end{equation}

\noindent where index $i_k$ corresponds to the distinct physical
levels of the $k$-th particle and $B^{[k],i_k}$ is a matrix with
elements $B^{[k],i_k}_{a_{k-1},a_k}$ such that the index $a_k$
forms a bond between the $k$-th particle and the $(k+1)$-th
particle, see Fig.~\ref{figure2}(a). The conventional rule for
tensor diagrams is that connected lines are summed over. We
additionally use arrows to indicate the matrix multiplication
order. Indices $a_0$ and $a_{n+1}$ are dummy and take the only
value, so $B^{[1],i_1}$ is a row matrix with elements
$B^{[1],i_1}_{1,a_1}$ and $B^{[n],i_{n}}$ is a column matrix with
elements $B^{[n],i_n}_{a_{n-1},1}$. $B^{[k]}$ is a rank-3 tensor
for $k=2,\ldots,n-1$ and a rank-2 tensor for $k=1,n$.
Fig.~\ref{figure2}(b) depicts a tensor representation for the
environment density operator, where we use complex conjugation
(denoted by $\ast$) to construct the bra-vector $\bra{\psi_E}$.
The partial trace over particles $k+1,\ldots,n$ results in a
tensor contraction shown in Fig.~\ref{figure2}(c). This
contraction becomes much simpler if we rewrite the MPS in the
\emph{right-canonical} form (which is always
possible~\cite{schollwock-2011,cirac-2021}), where
\begin{equation} \label{right-canonical-2}
\sum_{i_k=1}^d B^{[k],i_k} (B^{[k],i_k})^{\dag} = I_{k-1},
\end{equation}

\noindent with $I_{k-1}$ being the $|\{a_{k-1}\}| \times
|\{a_{k-1}\}|$ identity matrix. Then $\varrho_{1 \ldots k}$
entirely depends on tensors $B^{[1]},\ldots,B^{[k]}$, with the
irrelevant (future) particles $k+1,\ldots,n$ being replaced by a
single connecting line, see Fig.~\ref{figure2}(d).
Fig.~\ref{figure2}(d) contains an extra tensor $\chi_0$, which is
a trivial $1\times 1$ identity matrix in the case of a pure
environment. If the environment density operator is a mixture
$\varrho_E = \sum_q p_q \ket{\psi_E^q} \bra{\psi_E^q}$, where each
MPS $\ket{\psi_E^q}$ adopts the right-canonical form with matrices
$B^{[q,k],i_k}$, then $\chi_0 = {\rm diag}(p_1,p_2,\ldots)$ and
$B^{[k],i_k} = \bigoplus_{q} B^{[q,k],i_k}$. Therefore, the tensor
diagram for in Fig.~\ref{figure2}(d) is equally applicable to both
pure and mixed environment states. One could alternatively use the
formalism of matrix product density
operators~\cite{verstraete-2004,zwolak-2004} to represent the
mixed environment; however, this would not change the main idea
and would merely result in a slight modification of the Kraus
operators presented in Section~\ref{section-system-dynamics} (see
the review~\cite{filippov-2022} inspired by this paper).

The presented formalism is also applicable to the case when the
environment represents an infinite chain of particles in both
directions, e.g., the famous Affleck-Kennedy-Lieb-Tasaki (AKLT)
antiferromagnetic spin chain~\cite{aklt-1987}. The first collision
happens with some intermediate particle (the past particles are
assumed to be unaccessible). The partial trace over the past
particles results in the positive semidefinite matrix $\chi_0$
with unit trace. In all the scenarios, $\chi_0$ is a density
matrix for \emph{bond degrees for freedom}. To deal with the bond
degrees of freedom, we formally introduce an auxiliary Hilbert
space ${\cal H}_{\text{bond} \# k}$ spanning orthonormal vectors
$\{ \ket{a_k} \}$ as is shown in Fig.~\ref{figure2}(d). The matrix
$B^{[k],i_k}$ defines a mapping from ${\cal H}_{\text{bond} \# k}$
to ${\cal H}_{\text{bond} \# (k-1)}$ (from right to left in
Fig.~\ref{figure2}), whereas the transposed matrix
$(B^{[k],i_k})^{\top}$ defines a mapping from ${\cal
H}_{\text{bond} \# (k-1)}$ to ${\cal H}_{\text{bond} \# k}$ (from
left to right).

The maximum bond dimension $\max_k |\{a_k\}|$ (the MPS rank) for a
general state scales exponentially with the number of particles;
however, if the state is slightly entangled in terms of the
entanglement entropy [with potentially long correlations as in the
Greenberger-Horne-Zeilinger (GHZ) state], then such a state can be
efficiently described in the right-canonical form with a rather
small bond dimension~\cite{vidal-2003}. For instance, the MPS rank
equals $2$ for the GHZ state of $n$ qubits, the AKLT state of $n$
qutrits, the photonic cluster
state~\cite{istrati-2020,besse-2020}, and an arbitrary
single-photon wavepacket $\ket{\psi_E} = c_1 \ket{100 \ldots 00} +
c_2 \ket{010 \ldots 00} + \ldots + c_n \ket{000 \ldots 01}$. We
consider some of these states and a two-photon state from the
cascade emission with the MPS rank $3$ as examples in subsequent
sections.

\section{System dynamics}

\subsection{Markovian embedding} \label{section-system-dynamics}

\begin{figure}
\includegraphics[width=8.5cm]{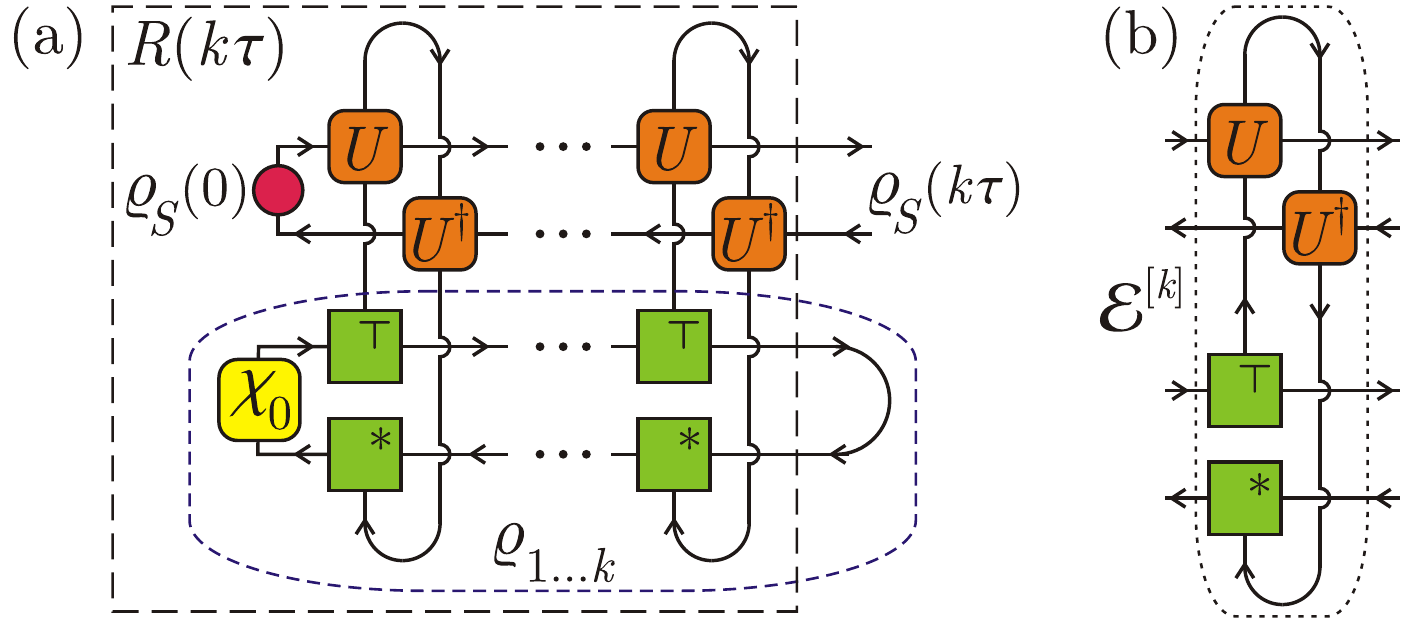}
\caption{(a) Tensor network diagram for the system density
operator $\varrho_S(k\tau)$ and the system-bond density operator
$R(k\tau)$ after $k$ collisions. (b) Completely positive and trace
preserving map ${\cal E}^{[k]}$.} \label{figure3}
\end{figure}

Were the environment uncorrelated, the system evolution would be
described by sequential applications of quantum channels
$\widetilde{\Phi}_k$ defined through
$\widetilde{\Phi}_k[\varrho_S] = {\rm tr}_k [U_{Sk} \varrho_{S}
\otimes \varrho_k U_{Sk}^{\dag}]$, where $\varrho_k$ is a density
operator for the $k$-th environment particle. As this is not the
case, we have to draw a full tensor diagram for collisions in
Fig.~\ref{figure3}(a). Upper $\cap$-lines correspond to the trace
over environment particles, which the system has already
interacted with. Looking at the diagram from left to right, we
observe the evolution of a rank-4 tensor $R(k\tau)$ that is a
composite \emph{system-bond density operator} on the Hilbert space
${\cal H}_S \otimes {\cal H}_{\text{bond} \# k}$ with $R(0) =
\varrho_S(0) \otimes \chi_0$. Due to the right-normalization
condition~\eqref{right-canonical-2}, the partial trace for
$R(k\tau)$ over bond degrees of freedom effectively produces the
reduced environment state $\varrho_{1 \ldots k}$ at the bottom of
the diagram and Eq.~\eqref{dynamics} yields the system density
operator,
\begin{equation}
\varrho_S(k\tau) = {\rm tr}_{\text{bond}\# k} [R(k\tau)].
\label{reduced-system-dynamics}
\end{equation}

\begin{figure}[b]
\includegraphics[width=8.5cm]{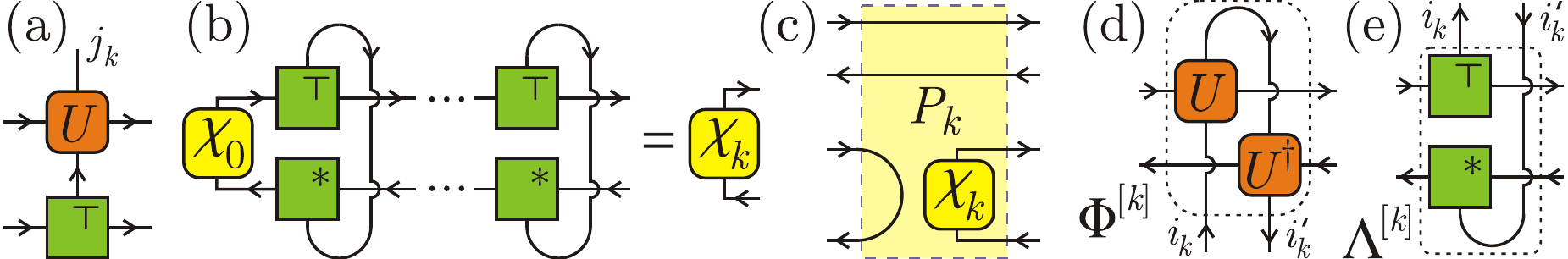}
\caption{Elementary tensor diagrams.} \label{figure4}
\end{figure}

These are the bond indices through which the information about the
previous collisions propagates in time and affects the system
evolution long time after the collisions actually happened. Time
evolution of the tensor $R$ is governed by unitary operators
$U_{Sk}$ as well as by tensors $(B^{[k],i_k})^{\top}$ that start
playing a role of evolution operators for the bond degrees of
freedom. The system-bond dynamics is given by a recurrent relation
\begin{equation}
R(k\tau) = {\cal E}^{[k]} \left[ R\big((k-1)\tau \big) \right],
\label{recurrent-relation-R}
\end{equation}

\noindent where a propagator map ${\cal E}^{[k]}$ is depicted in
Fig.~\ref{figure3}(b). This map is completely positive and trace
preserving due to the unitarity of $U_{Sk}$ and the
right-normalization condition~\eqref{right-canonical-2}. A
diagonal sum representation ${\cal E}^{[k]}[\bullet] = \sum_{j_k}
A_{j_k} \bullet A_{j_k}^{\dag}$ has the Kraus operators
\begin{equation} \label{Kraus}
A_{j_k} = \sum_{i_k} \bra{j_k} U_{Sk} \ket{i_k} \otimes
(B^{[k],i_k})^{\top}
\end{equation}

\noindent depicted in Fig.~\ref{figure4}(a).
Eqs.~\eqref{reduced-system-dynamics} and
\eqref{recurrent-relation-R} manifest the Markovian embedding for
the system dynamics. Such embeddings are of great use in
description of open quantum
systems~\cite{iles-smith-2016,tamascelli-2018,haase-2018,lu-2020,luchnikov-2020}.
Previous studies on Markovian embeddings for collision models
assumed no initial correlations in the
environment~\cite{kretschmer-2016,campbell-2018}. Our construction
is valid for a generally correlated MPS environment, with the MPS
rank being a dimension of an ``effective reservoir'' in the
embedding. A different but similar tensor network consideration of
an approximate Markovian embedding for a rather general open
system dynamics is reported in Ref.~\cite{luchnikov-2019}.

\subsection{Case study: Interaction with a two-photon wavepacket}

\begin{figure}[b]
\includegraphics[width=8.5cm]{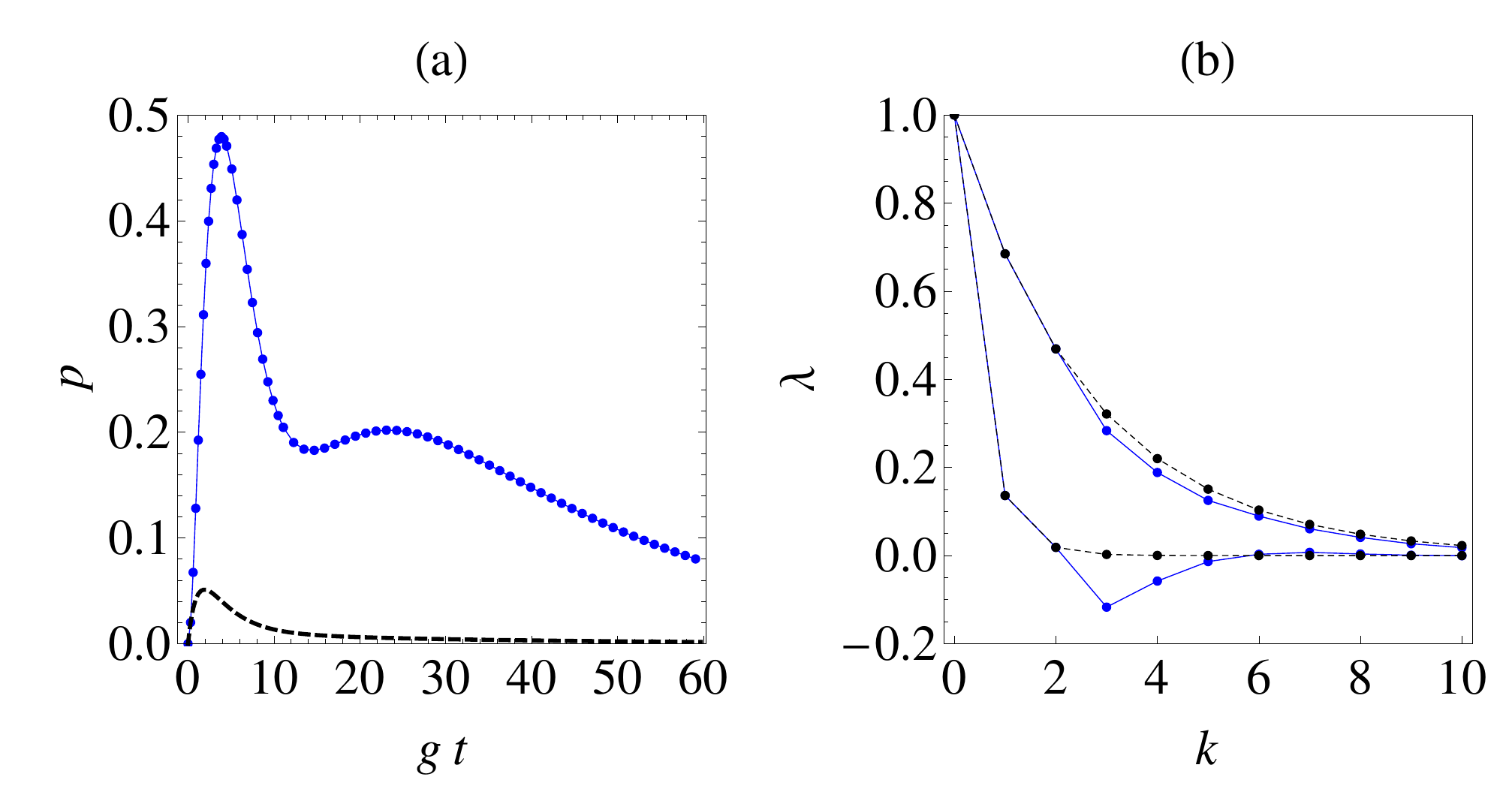}
\caption{(a) Excited level population vs dimensionless time for a
system interacting with a correlated two-photon wavepacket (upper
solid line). Disregard of correlations results in the lower dashed
line. Parameters $g\tau = 0.3$, $g T_1 = 2.3$, $g T_2 = 59.9$. (b)
Qubit coherence function vs number of system collisions with a
linear cluster state. Exact solution (blue solid lines) and
uncorrelated environment approximation (black dashed lines) for
parameters $g\tau = 0.3$ (upper two lines) and $g\tau = 0.6$
(lower two lines).} \label{figure5}
\end{figure}

Consider a two-level system in the ground state $\ket{g}$. The
system is exposed to a two-photon wavepacket, e.g., generated from
the cascade emissions~\cite{jen-2017,cere-2018}, with the time-bin
representation $\ket{\psi_E} \propto \sum_{l,m} e^{-l\tau/ T_1}
e^{-m\tau/ T_2} \ket{0 \ldots 0 1_l 0 \ldots 0 1_{l+m} 0 \ldots}$.
Each photon has an exponentially decaying temporal profile;
however, the second photon can only be emitted after the first
one. Such a wavepacket is a right-canonical MPS of rank $3$, where
$\chi_0 = {\rm diag}(1,0,0)$, $B^{[k],0} = {\rm diag}(e^{-\tau/
T_1},e^{-\tau/ T_2},1)$, and $B^{[k],1}$ has two non-zero elements
$B^{[k],1}_{a,a+1} = \sqrt{1-e^{-2\tau/ T_a}}$, $a=1,2$ for all
$k$~\cite{crosswhite-2008}. The energy levels of the system
interact with each time-bin mode via the excitation-preserving
exchange $U = \exp[g\tau(\ket{e}\bra{g} \otimes a^{\dag} -
\ket{g}\bra{e} \otimes a)]$, where $g$ has the physical dimension
of frequency, $a$ and $a^{\dag}$ are the photon annihilation and
creation operators, respectively. We treat $U$ as a $3 \times 3$
matrix because only $j_k = 0, 1, 2$ photons in each mode are
possible. The developed Markovian embedding theory enables us to
readily calculate the excited state population $p(t) = \bra{e}
\varrho_S(t) \ket{e}$, see Fig.~\ref{figure5}(a). The population
dynamics significantly differs from that for a factorized
radiation field $\bigotimes_{i} \varrho_i$, which illustrates the
strong effect of environment correlations on the system dynamics.

\subsection{Case study: Interaction with a photonic cluster state}

The environment state $\ket{\psi_E}$ is given by matrices $\chi_0
=
{\rm diag}(1,0)$, $B^{[k],0} = \frac{1}{\sqrt{2}} \left(%
\begin{array}{cc}
  1 & 0 \\
  1 & 0 \\
\end{array}%
\right)$, and $B^{[k],1} = \frac{1}{\sqrt{2}} \left(%
\begin{array}{cc}
  0 & 1 \\
  0 & -1 \\
\end{array}%
\right)$ for all $k$, which encode, e.g., photon-number
entanglement between modes~\cite{wein-2021}. Let a single
system-mode interaction be $U = \exp[g\tau (\ket{e}\bra{g} +
\ket{g}\bra{e}) \otimes (a - a^{\dag})]$.
Eqs.~\eqref{reduced-system-dynamics}--\eqref{Kraus}, where $j_k$
is now unlimited, result in a dephasing system dynamics with the
coherence basis states $(\ket{g} \pm \ket{e})/\sqrt{2}$ and the
decoherence function $\lambda$ shown in Fig.~\ref{figure5}(b). The
figure also depicts the coherence function if the environment
correlations are disregarded. Moreover, the first two collisions
result in the same dynamics for both correlated and uncorrelated
environments. A question arises: Why do correlated and
uncorrelated environments result in very different dynamics in
Fig.~\ref{figure5}(a) and very close dynamics in
Fig.~\ref{figure5}(b)? To anticipate the detailed analysis, which
we provide in what follows, the reason for that behavior is the
two-point environment correlation function, which significantly
differs from zero for the two-photon wavepacket and vanishes for
the cluster state. The small deviation in Fig.~\ref{figure5}(b) is
due to higher-order environment correlations.

\section{Master equation}

\subsection{Memory kernel and two-point correlations}

Though the Markovian embedding technique provides a universal
recipe for the system dynamics, the physics of dynamical memory
effects gets clearer in the time-convolution master equation,
\begin{equation} \label{nz-discrete}
\frac{\varrho_S \big( (k+1)\tau \big) - \varrho_S(k\tau)}{\tau} =
\sum_{m=0}^{k} {\cal K}_{km} [ \varrho_S \big((k-m)\tau \big) ],
\end{equation}

\noindent where the memory kernel map ${\cal K}_{km}$ relates the
density matrix increment with the past density operators. A
time-local term ${\cal K}_{k0}$ gives the density operator
increment caused by the latest collision (among those that have
already happened), whereas ${\cal K}_{km}$ for $m \geq 1$
describes a nontrivial effect of $m$ preceding collisions on the
system evolution. To derive the memory kernel we use the standard
projection operator techniques~\cite{breuer-2002} and adapt them
to our collision model. The main modification is in the
time-dependent nature of projection $P_k$ applied at time $k\tau$
to the system-bond density operator. We define $P_k [R] = {\rm
tr}_{\text{bond} \# k} [R] \otimes \chi_k$, where $\chi_k$ is a
bond density operator induced by a ``free evolution'' for the bond
degrees of freedom, i.e., $\chi_k = \sum_{i_k}
(B^{[k],i_k})^{\top} \chi_{k-1} (B^{[k],i_k})^{\ast}$, see
Figs.~\ref{figure4}(b,c). The projection $P_k$ breaks the
past-future correlations in the environment and yields $P_k [
R(k\tau) ] = \varrho_S(k\tau) \otimes \chi_k$. Inserting the
identity transformation ${\rm Id}_{S+\text{bond} \# k} = P_k +
Q_k$, where $Q_k$ is a complementary projection, in
Eq.~\eqref{recurrent-relation-R}, we solve a recurrent equation on
$Q_{k}[ R( k \tau ) ]$ with the initial condition $Q_0[R(0)] = 0$
and get an explicit solution for $P_{k+1}[ R \big( (k+1) \tau
\big) ]$, which yields the following kernel components: the local
term ${\cal K}_{k0} [\varrho_S] =
\frac{1}{\tau}(\widetilde{\Phi}_{k+1} [\varrho_S] - \varrho_S)$
and the nonlocal term ${\cal K}_{km} [\varrho_S] = \frac{1}{\tau}
{\rm tr}_{\text{bond} \#(k+1)} \circ {\cal E}^{[k+1]} \circ Q_k
\circ {\cal E}^{[k]} \circ \ldots \circ Q_{k-m+1} \circ {\cal
E}^{[k-m+1]} [\varrho_S \otimes \chi_{k-m}]$. The local term is
the only contribution to the memory kernel in the absence of
environment correlations.

To understand the nonlocal term, we decompose the embedding map
${\cal E}^{[k]} = \sum_{i_k,i'_k} \Phi^{[k]}_{i_k i'_k} \otimes
\Lambda^{[k]}_{i_k i'_k}$ into two parts, where only
$\Phi^{[k]}_{i_k i'_k}$ depends on the interaction nature, see
Figs.~\ref{figure4}(d,e). Then we find a series expansion for
$\Phi^{[k]}_{i_k i'_k}$ with respect to the interaction strength
$g \tau$ between the system and an individual environment
particle. Here we assume that the system-particle interaction
Hamiltonian during the $k$-th collision is $g \hbar H_k$, where
$\hbar$ is the reduced Planck constant and $H_k$ is a
dimensionless Hermitian operator with the operator norm $\|H_k\|
\leq 1$. A straightforward contraction of the tensor diagram for
${\cal K}_{km}$ yields the following largest contribution to the
memory kernel that comes from the second-order perturbation:
\begin{equation} \label{K-km-via-correlator}
{\cal K}_{km}^{(2)} [\varrho_S] = - g^2 \tau \left. {\cal C}_{ll'}
\left( \Big[ H_{l} , \big[H_{l'}, \varrho_S \big] \Big] \right)
\right\vert_{l=k+1,l'=k-m+1},
\end{equation}

\noindent where $[\cdot,\cdot]$ denotes the commutator and ${\cal
C}_{ll'} ( \bullet ) = {\rm tr}_{l,l'} [ \bullet (\varrho_{l,l'} -
\varrho_{l} \otimes \varrho_{l'}) ]$ is a two-point
operator-valued correlation function. For instance, ${\cal
C}_{ll'} (H_l \varrho_S H_{l'}) = \ave{H_l \varrho_S H_{l'}}_E -
\ave{H_l}_E \varrho_S \ave{H_{l'}}_E$.
Eq.~\eqref{K-km-via-correlator} provides an important physical
link between the environment correlation function and the memory
kernel.

\subsection{Stroboscopic limit}

If $\tau \ll 1/g$, then $\frac{1}{2\tau} [\varrho_S \big(
(k+1)\tau \big) - \varrho_S \big( (k-1)\tau \big)] =
\frac{d\varrho_S(t)}{dt} + O(g^3 \tau^2)$, where $t=k\tau$ is a
continuous time. If additionally the environment correlation
length $l_{\rm corr}$ is finite, then we can neglect the
contribution of $m$-point correlations ($m \geq 3$) in ${\cal
K}_{km}$ and get the celebrated Nakajima-Zwanzig
equation~\cite{nakajima-1958,zwanzig-1960}
$\frac{d\varrho_S(t)}{dt} = \int_0^t K(t')[\varrho_S(t-t')]dt'$
for a homogeneous collision model, where $U_{Sk}$, $B^{[k]}$, and
$\chi_k$ do not depend on $k$. The kernel $K(t')[\varrho_S] =
\delta(t') L_{\rm local}[\varrho_S] + \frac{1}{2} g^2 \tau
\sum_{m=1}^{\infty} \delta(t'-m\tau) K_m[\varrho_S] +
O(g^3\tau^2)$, where $\delta$ is the Dirac delta function, $L =
\frac{1}{2\tau}(\widetilde{\Phi}_{12}-{\rm Id}_S)$ originates from
two sequential collisions, and $K_m[\varrho_S] = \big[ \ave{H}_E ,
[ \ave{H}_E , \varrho_S ] \big] - \left\langle \big[ H_{m+1} ,
[H_{1}, \varrho_S \otimes I_E ] \big] \right\rangle_E$ describes
the exponentially decaying correlations in an
MPS~\cite{perez-garcia-2007,verstraete-2008,schollwock-2011,cirac-2021},
so $K_m[\varrho_S] = (\pm 1)^m e^{-m/l_{\rm corr}} L_{\rm
nonlocal}[\varrho_S]$, where $\pm$ is a sign of the second largest
eigenvalue of the transfer matrix. The kernel $K(t')$ is the
inverse Laplace transform~\cite{chruscinski-2010,smirne-2010} of
$(\pm e^{s \tau + l_{\rm corr}^{-1}} - 1)^{-1} L_{\rm nonlocal}$.
In the stroboscopic limit $g\tau \rightarrow 0$, $g^2\tau = {\rm
const}$, which is discussed in
Refs.~\cite{giovannetti-2012,luchnikov-2017,lorenzo-2017}, we get
the exact equation $\frac{d\varrho_S(t)}{dt} = L[\varrho_S(t)]$ of
the Gorini-Kossakowski-Sudarshan-Lindblad
form~\cite{gks-1976,lindblad-1976} with $L = L_{\rm local} +
\frac{1}{2} g^2 \tau (\pm e^{l_{\rm corr}^{-1}} - 1)^{-1} L_{\rm
nonlocal}$. Importantly, the relaxation rate in $L$ may
significantly differ from that in $L_{\rm local}$. Higher order
stroboscopic limits are discussed in more detail in the
review~\cite{filippov-2022} inspired by this paper.

\begin{figure}
\includegraphics[width=8.5cm]{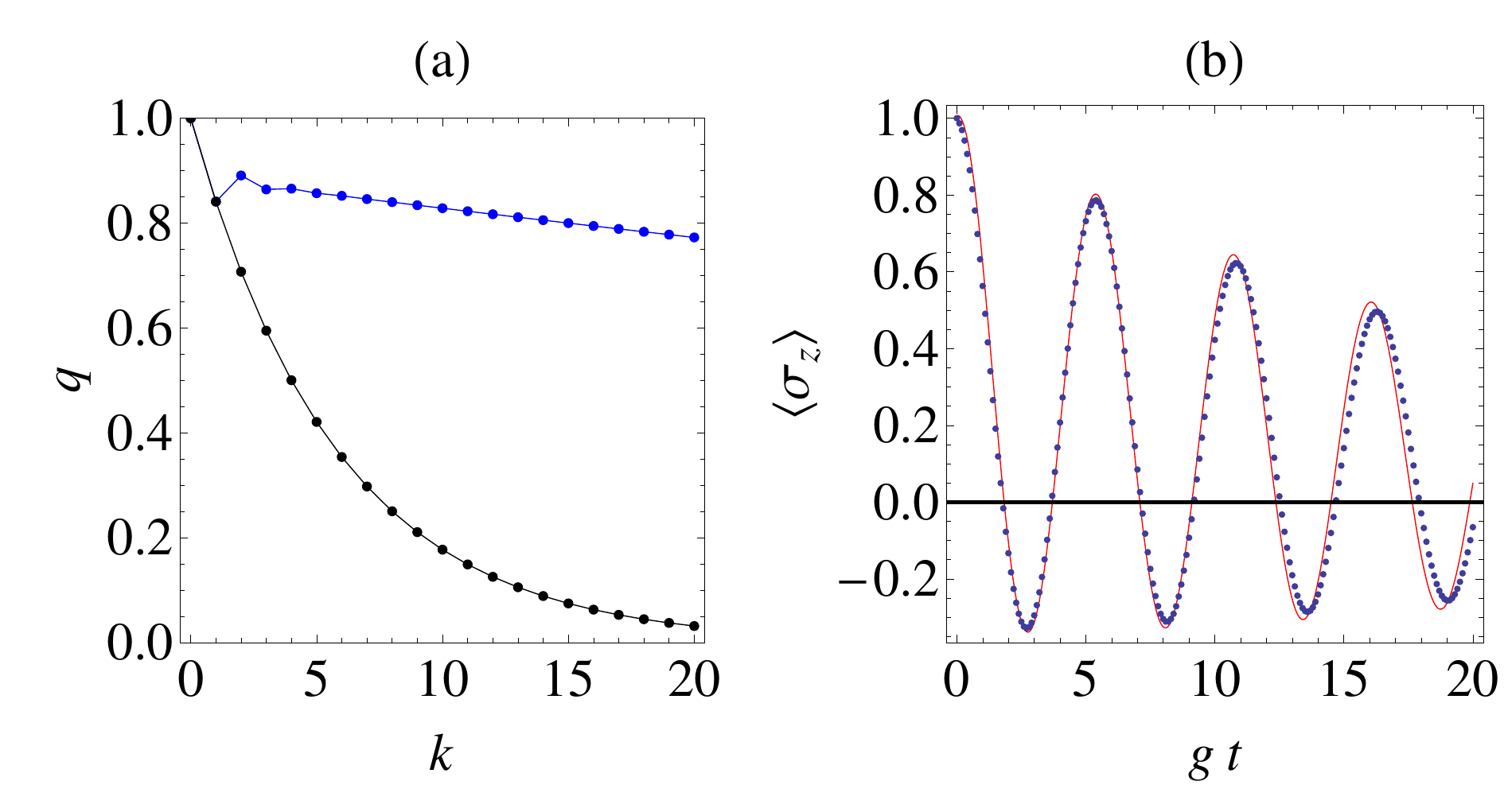}
\caption{(a) Qubit depolarization parameter vs number of
Heisenberg-interaction collisions with the AKLT spin chain: exact
solution (upper line) and uncorrelated environment assumption
(lower line). (b) Qubit observable vs dimensionless time in an
exemplary collisional dynamics with the AKLT environment: exact
(dots) and stroboscopic limit (solid line).} \label{figure6}
\end{figure}

\subsection{Case study: Interaction with AKLT infinite spin chain}

The AKLT state of spin-1 particles is a right-canonical MPS of
rank $2$ with matrices $B^{[k],0} = {\rm diag}(-1 / \sqrt{3}, 1 /
\sqrt{3})$ and $B^{[k],\pm 1}$ that have the only nonzero element
$B^{[k],1}_{12} = - B^{[k],-1}_{21} = \sqrt{2 / 3}$. At time $t=0$
a qubit system collides with one of the chain spins, then collides
with its right neighbor and so on. In this scenario, $\chi_0 =
\frac{1}{2} I$. Consider the Heisenberg-type qubit-spin
interaction $U = \exp[-\frac{g \hbar}{2} ( \sigma_x \otimes J_x +
\sigma_y \otimes J_y + \sigma_z \otimes J_z ) ]$, where
$(\sigma_x,\sigma_y,\sigma_z)$ is the set of Pauli matrices and
$J_{\alpha}$ is an operator for the spin projection (in units of
$\hbar$) on the $\alpha$ direction. The AKLT state has
exponentially decaying two-point correlations because
$\varrho_{1m} = \frac{1}{3} I \otimes \frac{1}{3} I + \left(
-\frac{1}{3} \right)^{m} \left( J_x \otimes J_x + J_y \otimes J_y
+ J_z \otimes J_z \right)$; however, these correlations are strong
enough to significantly deviate the qubit dynamics from that for
the uncorrelated environment.  The disregard of environment
correlations yields the qubit dynamics $\varrho_S(t) = q(t)
\varrho_S(0) + [1-q(t)] \frac{1}{2} I$, where the depolarization
function $q_{\rm Markov}(k\tau) = [\frac{1}{27} \left( 11 + 16
\cos \frac{3}{2} g \tau \right)]^k$ has the asymptotic behavior
$q_{\rm Markov}(t) \approx \exp(- \frac{2}{3} g^2 \tau t)$ if
$g\tau \ll 1$. However, the exact qubit dynamics is given by
$q(k\tau) = \left( \frac{1}{2} + \frac{x}{z} \right) \left(
\frac{y + z}{27} \right)^{k} + \left( \frac{1}{2} - \frac{x}{z}
\right) \left( \frac{y - z}{27} \right)^{k}$, where $x = 2 + 7
\cos \frac{3 g \tau}{2}$, $y = 7 + 2 \cos \frac{3}{2} g \tau$, $z
= 2 \sqrt{y^2 + 27 \sin^2 \frac{3}{2} g\tau}$. Hence, $q(t)
\approx (1-\frac{1}{2} g^2 \tau ^2)\exp(-\frac{1}{8} g^4 \tau^3
t)$ if $g\tau \ll 1$, see Fig.~\ref{figure6}(a). The exponent
power vanishes in the stroboscopic limit, so does $L$. To
demonstrate efficacy of the stroboscopic-limit equation
$\frac{d\varrho_S(t)}{dt} = L[\varrho_S(t)]$ with nonvanishing
decoherence rate, we consider a controlled unitary interaction $U
= e^{-ig\tau\sigma_x} \otimes \ket{+1}\bra{+1} +
e^{-ig\tau\sigma_y} \otimes \ket{0}\bra{0} + e^{-ig\tau\sigma_z}
\otimes \ket{-1}\bra{-1}$ with $g\tau = 0.1$ and show a good
agreement between the exact and approximate dynamics in
Fig.~\ref{figure6}(b). These examples illustrate that the
two-point environment correlations correctly describe the system
dynamics under the stroboscopic assumption $g \tau \ll 1$ if
$l_{\rm corr}$ is finite. If $l_{\rm corr} = \infty$ (e.g., for
the GHZ state), then multitime correlation functions are to be
taken into consideration too.

\section{Conclusions}

We have presented two approaches to the collisional open quantum
dynamics with a \emph{generally} correlated environment: the
Markovian embedding in
Eqs.~\eqref{reduced-system-dynamics}--\eqref{recurrent-relation-R}
and the time-convolution master equation~\eqref{nz-discrete} with
its continuous limit. The former approach readily provides a
solution to a number of timely problems like the system
interaction with two-photon wavepackets, structured photonic
states, and one-dimensional spin chains. The latter approach
reveals the physics of memory effects and its relation to the
environment correlation functions. Here we have demonstrated the
advantages of tensor networks in general collisional dynamics,
thus extending the range of successful tensor-network applications
in many-body dynamics~\cite{banuls-2009,manzoni-2017,lerose-2021},
operational meaning of
non-Markovianity~\cite{pollock-pra-2018,pollock-prl-2018,white-2020},
and spin-boson
models~\cite{strathearn-2018,jorgensen-2019,gribben-2022}.

\end{document}